# A Comparison of Measured and Predicted Wave-Impact Pressures from Breaking and Non-breaking Waves

A. M. Fullerton, T. C. Fu, and S. Brewton
Naval Surface Warfare Center, Carderock Division, USA

K.A. Brucker, T.T. O'Shea, and D.G. Dommermuth
Science Applications International Corporation, USA

**ABSTRACT**

Impact loads from waves on vessels and coastal structures are complex and may involve wave breaking, which has made these loads difficult to estimate numerically or empirically. Results from previous experiments have shown a wide range of forces and pressures measured from breaking and non-breaking waves, with no clear trend between wave characteristics and the localized forces and pressures that they generate. In 2008, a canonical breaking wave impact data set was obtained at the Naval Surface Warfare Center, Carderock Division, by measuring the distribution of impact pressures of incident non-breaking and breaking waves on one face of a cube. This experimental effort was sponsored by the Office of Naval Research (ONR), under the Dynamics of Interacting Platforms Program, Program Manager Dr. Ron Joslin. The effects of wave height, wavelength, face orientation, face angle, and submergence depth were investigated. Additionally, a limited number of runs were made at low forward speeds, ranging from about 0.5 to 2 knots (0.26 to 1.03 m/s).

The measurement cube was outfitted with a removable instrumented plate measuring 0.09 m$^2$ (1 ft$^2$), and the wave heights tested ranged from 20.3 to 35.6 cm (8-14 inches). Additionally, focused breaking waves were generated using a combination of non-breaking waves of various frequencies. The instrumented plate had 9 slam panels of varying sizes made from polyvinyl chloride (PVC) and 11 pressure gages; this data was collected at 5 kHz to capture the dynamic response of the gages and panels and fully resolve the shapes of the impacts. Other measurements included velocity, free surface, and video. Impact loads on the plate tended to increase with wave height, as well as with plate inclination toward incoming waves.

Numerical predictions of this experiment have been made using an ONR supported computational fluid dynamics code, Numerical Flow Analysis (NFA). Still images from the output of this code show similar phenomena to the experiment, as captured by video sequence. NFA was used previously to generate the same breaking wave produced in the experiment (Fu, et. al., (2008)). The test cube has now been added to the model, and pressures and velocities are predicted and compared to the experimental data. Peak predicted loads for the breaking wave cases are in the same range as the average peak measured pressures for the same conditions. Further comparisons of the experimental results with predicted values of the pressures and forces, and how they vary with wave characteristics and cube draft are investigated and presented in this paper.

**INTRODUCTION**

The magnitude of wave impact loads varies greatly, depending upon whether the wave is breaking, as well as on the wave height, length, steepness, and the geometry and immersion of the impacted structure. Chan and Melville (1984, 1987, 1988, & 1989) have performed several experiments to investigate the force of plunging breakers on flat plates and vertical cylinders. The results of these investigations show breaking wave impact pressures as high as $10\rho c^2$ on plates and cylinders, where ρ is the density of water and c is the wave celerity, which correspond to almost 96 kilopascals (2000 pounds per square foot (psf)) for a wavelength of 6.1 m (20 feet). Experimental results from Zhou, Chan and Melville (1991) found breaking wave impact pressures of up to $15\rho c^2$ on a vertical cylinder, which correspond to 144 kilopascals (3000 psf) for a wavelength of 6.1 m (20 ft). Field data collected by Bullock and Obhrai (2001) shows pressures of over 383 kilopascals (8000 psf) on a breakwater for an incident wave height of about 3.1 m (10 feet). These results suggest that there can be variation in the magnitudes of incident wave loads, and that wave impact pressures are dependent on wave characteristics.

This paper describes an experiment that was performed to characterize the distribution of breaking and non-breaking wave impact loads over a flat surface, similar to those performed in 2005 (Pence, et al. (2006), Hess, et al. (2006), Fullerton, et al. (2007) with non-breaking wave impact loads and those performed in 2007 (Fu, et al. (2008)) with breaking wave impact loads. In those experiments, the average loads were measured on a flat plate and a cylinder. In order to better understand the distribution of forces over a surface, the impact pressures in this experiment were measured on an instrumented test cube by using an array of slam panels and pressure gages. The objective of this work was to develop an improved understanding of the physics of breaking wave impacts, and to investigate the trends of wave slap loads under various wave height, wavelength, impact angle, and draft conditions. Results from this experiment are compared with computational results from Numerical Flow Analysis (NFA).

The NFA code provides turnkey capabilities to model breaking waves around a ship and ocean structures, including both plunging and spilling breaking waves, the formation of spray, and the entrainment of air. A description of the NFA and its current capabilities can be found in Dommermuth, et al. (2007); O'Shea, et al. (2008); and Brucker, et al. (2010). NFA solves the Navier-Stokes equations utilizing a Cartesian-grid formulation. The flow in the air and water is modeled. As a result, NFA can directly model air entrainment and the generation of droplets. The interface capturing of the free surface uses a second-order accurate, volume-of-fluid technique. A cut-cell method is used to enforce free-slip boundary conditions on the body. A boundary-layer model has been developed (Rottman, Brucker, Dommermuth & Broutman (2010)), but it is not used in these numerical simulations. NFA uses an implicit Subgrid-scale model that is built into the treatment of the convective terms in the momentum equations (Brucker, et al. (2010)). A surface representation of the ship hull (body) is all that is required as input in terms of body geometry. The numerical scheme is implemented on parallel computers using Fortran 90 and MPI. Relative to methods that use a body-fitted grid, the potential advantages of NFA's approach are significantly simplified gridding requirements and greatly improved numerical stability due to the highly structured grid.

**EXPERIMENTAL DESCRIPTION**

Wave impact testing was completed in August and September 2008. The instrumented cube and side boxes were suspended from Carriage 5 and held stationary in the High Speed Tow Basin at the Naval Surface Warfare Center, Carderock Division (NSWCCD), approximately 61 m (200 ft) from the wavemaker. The total length of the basin is approximately 514 m (1687 feet), and the width of the basin in this section is 6.4m (21 ft). This basin is equipped with a pneumatic wavemaker dome which is connected to a blower powered by a direct coupled variable speed DC electric motor rated at 75 kW (100hp), 1150 rpm.

The distribution of wave impact loads was measured over the top and front faces of a 0.305 cubic meter (one cubic foot) model constructed of aluminum. The measurement cube was outfitted with a removable instrumented plate which had 9 slam panels of varying sizes and 11 pressure gages (Figure 1). Two side boxes were attached to either end of the model to limit the wave effects to two dimensions (Figure 2).

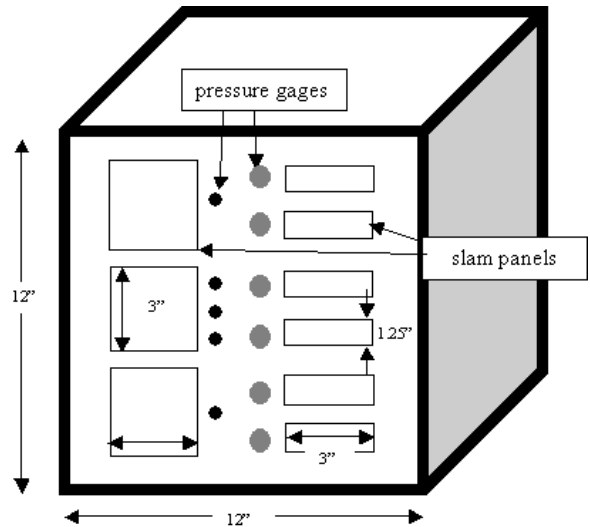

**Figure 1:** Diagram of test cube with slam panels and pressure gages.

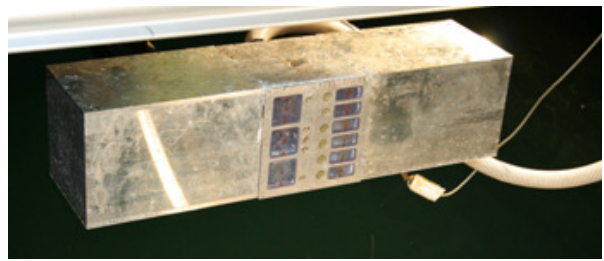

**Figure 2:** Instrumented cube.

Table 1 shows the test matrix for this experiment. Three different non-breaking conditions and one breaking wave condition were tested. Tests were run at three different levels of cube submergence (full, half and none), and three different cube angles (0°, +45° toward the incoming wave, -45° away from the incoming wave), which are shown in Figure 3. The instrumented plate was used on both the front and top face of the cube. Limited runs were made with forward speed in order to investigate the added forces due to forward motion, over a range of 0.5 to 2 knots (0.26 to 1.03 m/s). Runs were typically made for 3-5 minutes to gather sufficient wave data, which allowed for approximately 90-150 waves to pass (wave periods ranged from 1.97 s to 2.4 s). Each condition was run twice to examine repeatability. The stationary, front face, 0° cube angle impact results will be discussed in this paper.

The experimental breaking waves were generated by sending an external voltage signal made up of 9 waves of varying frequencies (Figure 4), using a method similar to previous experiments (Fu, et al. (2008)). The shortest waves were sent first, with increasingly longer waves being sent out in sequence. Since the speed of an individual wave is proportional to the square root of its wavelength, a shorter wave will travel more slowly than a longer wave, and all the waves will meet at some prescribed distance from the wavemaker. These individual waves were chosen to combine approximately 61m (200 ft) from the wavemaker to create a breaking wave. The waves that were generated to create the breaking wave are the larger waves in Figure 4 (greater than 6 volts); the smaller waves are inserted to create a smooth input signal so that the wavemaker did not have to come to an abrupt stop between waves. The breaking wave was created using the voltage input shown with a blower speed of 1600 RPM; a wave measurement near the plate shows the typical shape of the resultant wave, shown in Figure 5. Regular waves were generated through the specification of blower speed (RPM) and frequency (Hz).

**Table 1:** Test matrix for cube wave impact test.

| Wave Height | | Wave Length | | Breaking? | Face | Cube Angle | Plate Draft | Speed |
|---|---|---|---|---|---|---|---|---|
| (in) | (cm) | (ft) | (m) | | | (degrees) | | (kts) |
| 8,12 | 20.3,30.5 | 20 | 6.1 | no | front | 0,+45,-45 | none,half,full | 0 |
| 14 | 35.6 | 30 | 9.1 | no | front | 0,+45,-45 | none,half,full | 0 |
| 10 | 35.6 | n/a | n/a | yes | front | 0,+45,-45 | none,half,full | 0 |
| 8,12 | 20.3,30.5 | 20 | 6.1 | no | top | 0 | half, full | 0 |
| 14 | 35.6 | 30 | 9.1 | no | top | 0 | half, full | 0 |
| 10 | 35.6 | n/a | n/a | yes | top | 0 | none,half,full | 0 |
| 14 | 35.6 | 30 | 6.1 | no | front | 0 | none | 0.5,1,2 |
| 10 | 35.6 | n/a | n/a | yes | front | 0 | none | 0.5 |

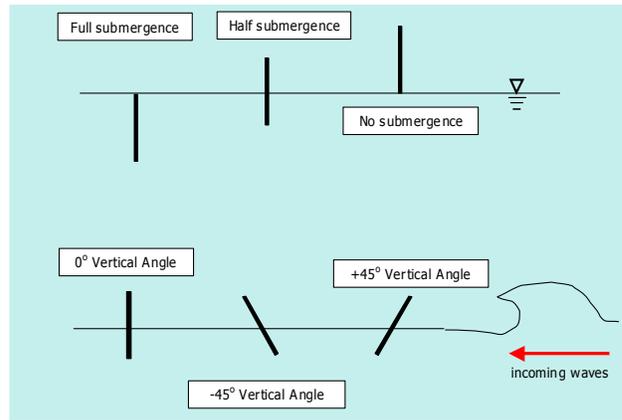

**Figure 3:** Instrumented plate submergence levels and angles.

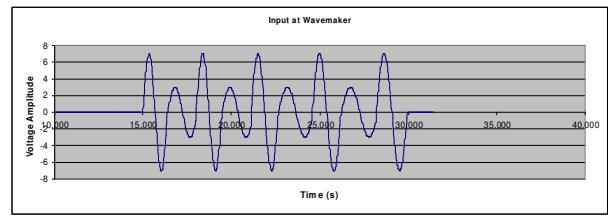

**Figure 4:** Wavemaker voltage input for breaking wave.

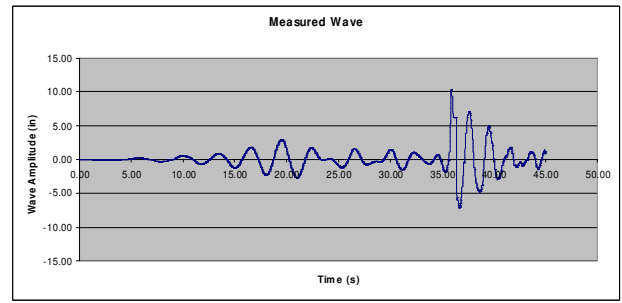

**Figure 5:** Wave measurement near plate.

## INSTRUMENTATION

### Slam panels

Nine slam panels were used on the instrumented face of the cube. Figure 6 shows the layout of these panels. The panels were made from rigid polyvinyl chloride (PVC). A standard panel thickness of approximately 0.245 cm (0.1 inches) was used. Each panel was instrumented with two strain gages wired into a Wheatstone bridge, which produced an output voltage proportional to the differential bending of the panel. The panels were calibrated to a uniform pressure measurement over their area, which was performed by submerging the cube and relating the panel response to the pressure gage readings. The panels were set to collect pressure samples at a rate of 5 kHz.

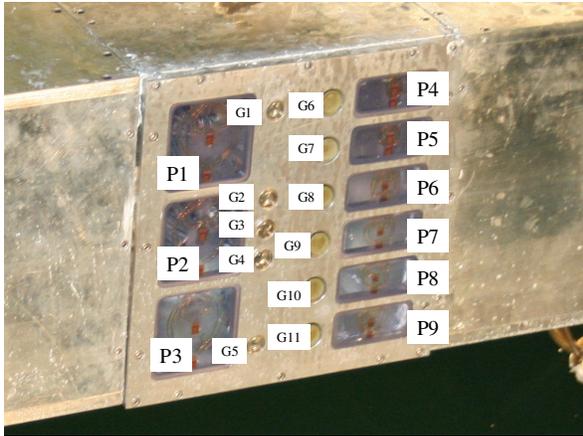

**Figure 6:** Numbering of panels and pressure gages on cube face.

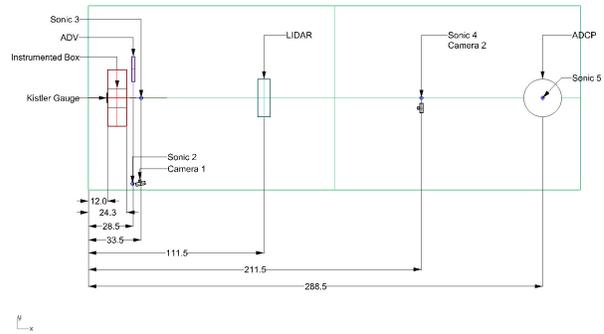

**Figure 7:** Plan view of experimental layout. All measurements are in inches.

**Pressure gages**

Eleven pressure gages were used on the instrumented face of the cube model. Six of the sensors that were used were the GE Novasensor NPI-19B-015AV, which can measure up to 103 kilopascals (15 pounds per square inch (psi), shown as gray circles on the right in Figure 2, G6-G11 in Figure 6). The GE Novasensor has a piezoresistive sensor chip housed in a fluid-filled cylindrical cavity which is isolated from the measured media by a stainless steel diaphragm and body, minimizing the temperature sensitivity of the gages. These gages have threaded ports which were fitted with a water-filled insert to prevent air from being trapped in the port, which might result in incorrect pressure readings. The other five pressure sensors were the GE Novasensor NPI-19A-015AV (shown as black circles on the left in Figure 1, G1-G5 in Figure 6). These are the same sensors as the other six, except that they have no port. All pressure gages were calibrated to 34.5 kilopascals (5 psi) using air pressure, and gages were set to collect pressure samples at rate of 5 kHz.

**Kistler gage**

Integrated force and moment measurements were made using a Kistler Gage. The Kistler was calibrated for a maximum load of 100 lbs (444.8 Newtons). The Kistler gage was mounted between the cube and support structure, and was set to collect pressure samples at rate of 5 kHz.

**Senix Ultrasonic Sensors**

The incoming waves were measured utilizing Senix ultrasonic distance sensors, which are non-contact, ultrasonic instruments for measuring distances in air. Five sensors were used, with three in line with the incoming waves: one far forward of the impact region to measure the incoming wave, one just in front of the model to measure the impacting/reflected wave, and one in between. Another sensor was aligned with the one just in front of the model, but was offset from the plate to measure the wave height in absence of the model. The final sensor was located downstream of the cube to measure the wave height after impact. Figure 7 shows a sketch of the experimental layout, which includes the ultrasonic sensor locations labelled Sonic #2 through Sonic #5. Sonic #1 is outside the range of this schematic.

**COMPUTATIONAL DESCRIPTION**

The parameters in the cube impact experiments described both above and in Fullerton, et al. (2009) were used as the basis for three-dimensional numerical simulations. In the normalization employed, $L_0=1/k$, where k is the wave number, $U_0=(g/k)^{1/2}$ where g is the acceleration due to gravity and $T_0=(kg)^{-1/2}$. The length, width, depth, and height of the computational domain was respectively $8\pi$ x $\pi/10$ x $2\pi$. The domain was discretized with 2048x128x768= 201,326,592 grid points, and it was distributed over 128x64x128 blocks and 192 cores. Inflow and outflow boundary conditions were used in the stream-wise, periodic in the span-wise, and free-slip in the cross-stream directions. The grid was close to uniform near the cube where the grid spacing was [0.003, 0.0025, 0.0031] in the x,y,z directions respectively. The maximum grid spacing far away from the cube was [0.015, 0.0025, 0.025]. In the breaking wave cases two time steps were used. For t<4.25, Δt=0.0005 and for t>4.25, Δt=0.00025. In the regular wave cases Δt=0.0005. All simulations were run on the SGI Altix ICE machine Diamond located at the ERDC. The simulations required between 30,000 and 50,000 iterations each. This required between 100-160 hours of wall-clock time and 20,000-33,000 CPU hours per simulation. Pressure, velocity and free-surface elevation data was output every 20 iterations for the breaking wave cases and every 100 iterations for the non-breaking wave cases. This data output rate corresponds to a dimensional sampling frequency of 636Hz and 318Hz, respectively.

The simulations, similar to the experiments, required either a single or multiple modes to generate a non-breaking or breaking wave with a wave height to length ratio of 0.06. The simulations differ from the experiments in that they use the pressure forcing technique described and validated in Dommermuth, et al. 2010 and Brucker, et al. 2010 to generate the waves. Here, the forcing period, $T_f$, is $2\pi$, the offset time $T_u$, is 8.0, $A_f=U_c=0$, and the amplitudes, wavenumbers, and frequencies of the modes are listed in Table 2.

**Table 2:** Parameters used with the atmospheric forcing technique of Brucker et al. 2010 to generate waves in the NFA simulations. Length and time scales are respectively, $L_0=1/k$ and $T_0=(kg)^{-1/2}$.

| Mode | Wavenumber | Frequency | Amplitude |
|---|---|---|---|
| Breaking | | | |
| 1 | 1.0 | 1.0 | 0.0594 |
| 2 | 2.0 | 1.4142 | 0.0297 |
| 3 | 3.0 | 1.7321 | 0.0150 |
| 4 | 4.0 | 2.0 | 0.0113 |
| Non-breaking | | | |
| 1 | 1.0 | 1.0 | 0.1 |

**RESULTS AND COMPARISONS**

Figure 8 shows the comparison between experimental and numerical results for the non-breaking 30.5 cm height wave condition. The comparison is of the free-surface elevation 22.86 cm (9 in) in front of the cube. The NFA computations for regular waves generated a regular wave with a height of 35.6 cm (14 in) and length of 6.1 m (20 ft). Though this is different than the desired wave from the experiment which had a height of 30.5 cm (12 in) for a length of 6.1 m (20 ft), the actual height of the individual regular waves varied over an experimental run, so a single wave with similar characteristics was chosen from the several regular impacts measured for the 30.5 cm (12 inch) desired height condition. The wave time series compare well, with the first numerical wave being slightly smaller than the second due to start up transients. Details of the numerical procedure that is used to generate waves is provided in Brucker, et al., 2010. Figure 9 shows the comparison between experimental and numerical results for the breaking wave condition. Again, the wave time series compare well. The NFA computations were designed to generate a breaking wave which broke very near the front face of the cube with a value of h/λ near that of the experiments.

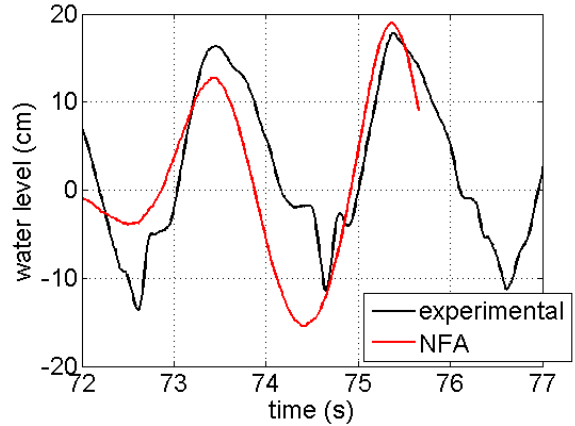

**Figure 8:** Comparison of non-breaking wave time series just in front of the cube (at the Sonic 3 location from Figure 7).

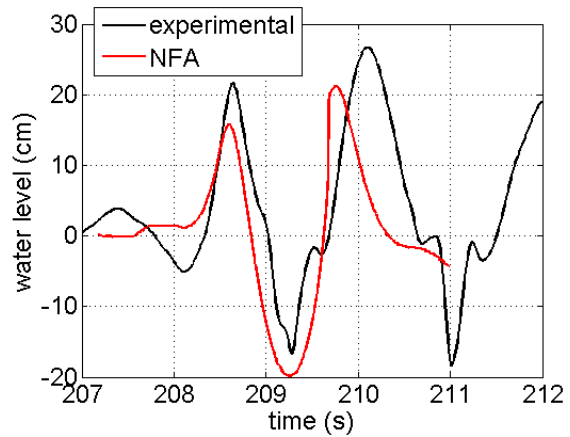

**Figure 9:** Comparison of breaking wave time series just in front of the cube (at the Sonic 3 location from Figure 7).

Figures 10 through Figure 15 show comparisons of the experimental and numerical results for the non-breaking wave impacts for the wave conditions specified in Figure 8 for the case of 0° cube angle and three different submergence levels. Comparisons are made at two pressure measurement levels: 18.10 cm (7.13 in) above the bottom of the cube (G2, Figure 6) and 4.13 cm (1.63 in) above the bottom of the cube (G11, Figure 6). Experimental results are shown in black and numerical results are shown in red. The pressures reference the calm water pressure condition as zero. A single experimental impact is shown that appeared to match the prediction, however impact pressures can vary over the same condition in the experiment, so the experimental average, maximum and minimum values are also plotted (blue symbols). These comparisons show that the predicted values are within the experimental range of values, and that the shape of the peak is similar between experiments and numerics. The simulation results are averaged over the span-wise direction at the probe heights. The span-wise averaging would tend to eliminate local minimum and maximum values.

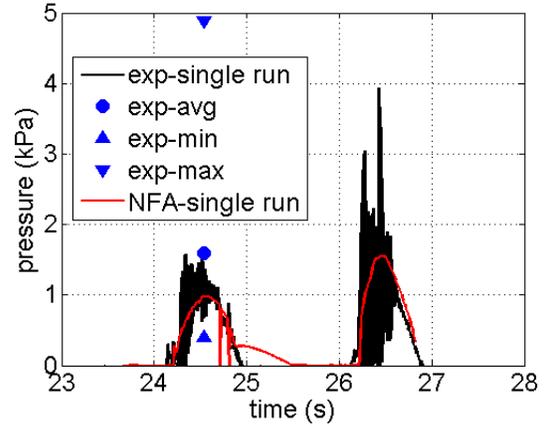

**Figure 11:** Comparison of experimental results for non-breaking wave (30.5 cm height) from G11 and computational results at 4.13 cm (1.63 in) above the bottom of the cube with 0° cube angle and no submergence.

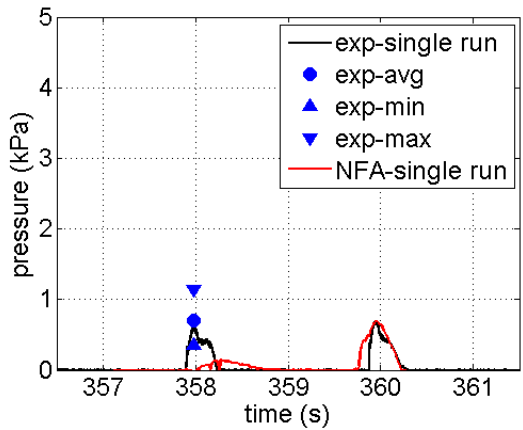

**Figure 10:** Comparison of experimental results for non-breaking wave (30.5 cm height) from G2 and computational results at 18.10 cm (7.13 in) above the bottom of the cube with 0° cube angle and no submergence.

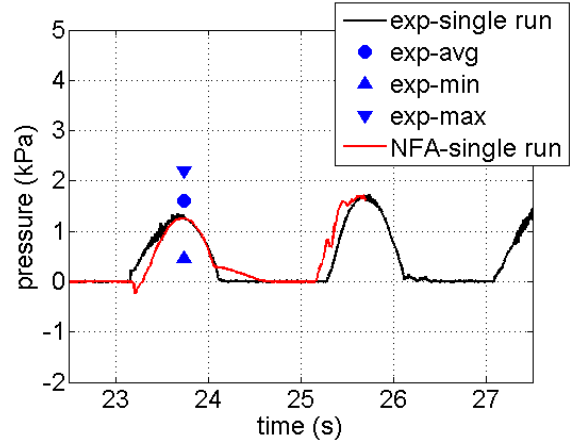

**Figure 12:** Comparison of experimental results for non-breaking wave (30.5 cm height) from G2 and computational results at 18.10 cm (7.13 in) above the bottom of the cube with 0° cube angle and half submergence.

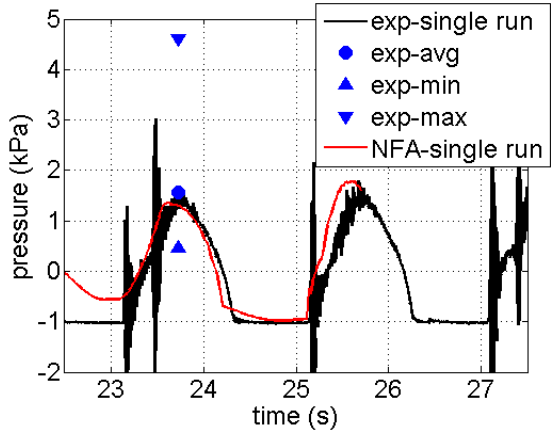

**Figure 13:** Comparison of experimental results for non-breaking wave (30.5 cm height) from G11 and computational results at 4.13 cm (1.63 in) above the bottom of the cube with 0° cube angle and half submergence.

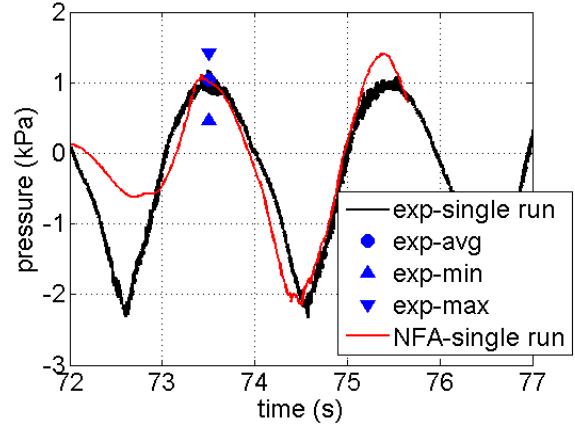

**Figure 15:** Comparison of experimental results for non-breaking wave (30.5 cm height) from G11 and computational results at 4.13 cm (1.63 in) above the bottom of the cube with 0° cube angle and full submergence.

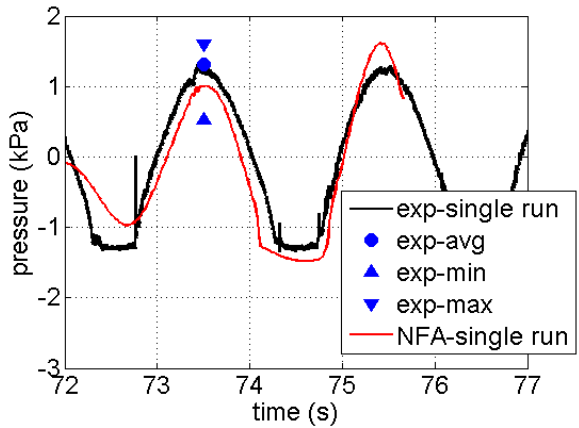

**Figure 14:** Comparison of experimental results for non-breaking wave (30.5 cm height) from G2 and computational results at 18.10 cm (7.13 in) above the bottom of the cube with 0° cube angle and full submergence.

Figure 16 shows an image series from the experimental breaking wave case of no submergence and 0° cube angle, and Figure 17 shows an instantaneous image from NFA for the same case. Both of these figures show the nature of the breaking wave cases with the wave breaking into droplets, which causes great variation in the measured pressures over the cube face.

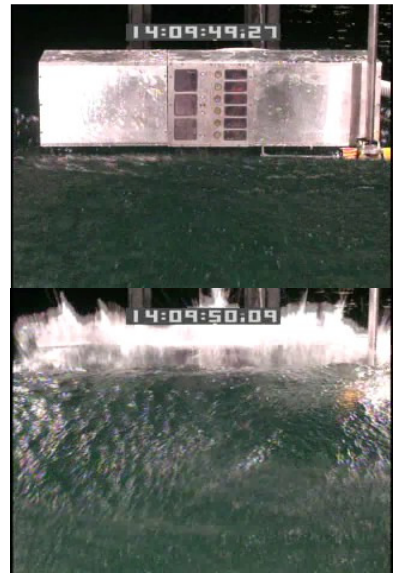

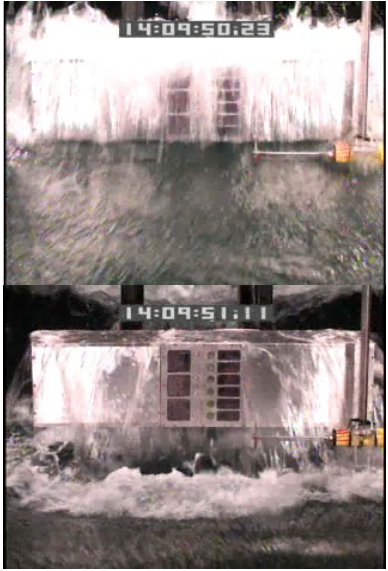

**Figure 16:** Image series from the experiment for the breaking wave case on cube with no submergence.

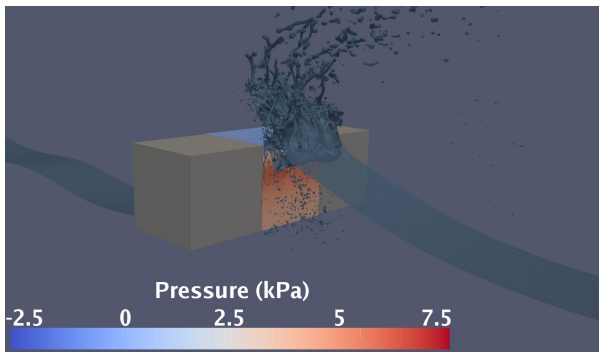

**Figure 17:** Instantaneous image from NFA for breaking wave on cube with no submergence.

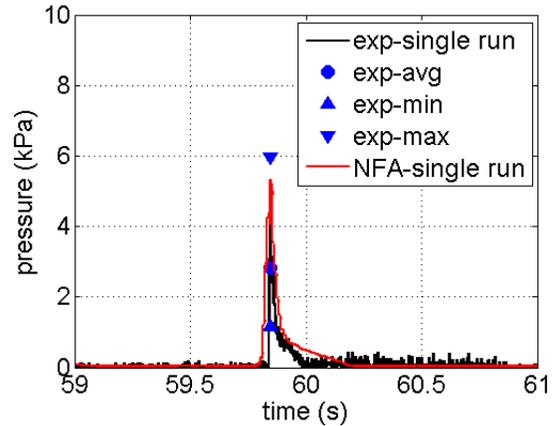

**Figure 18:** Comparison of experimental results for breaking wave from G2 and computational results at 18.10 cm (7.13 in) above the bottom of the cube with 0° cube angle and no submergence.

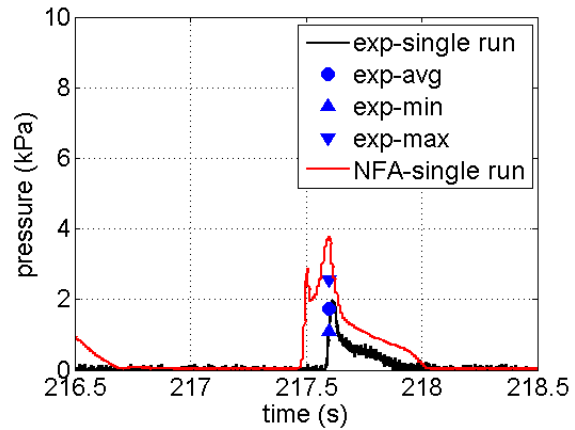

**Figure 19:** Comparison of experimental results for breaking wave from G11 and computational results at 4.13 cm (1.63 in) above the bottom of the cube with 0° cube angle and no submergence.

Figure 18 through Figure 23 show comparisons of the experimental and numerical results for the breaking wave cases. Again, experimental results are shown in black and numerical results are shown in red. The pressures reference the calm water pressure condition as zero. A single experimental impact is shown, however impact pressures can vary greatly over the same breaking condition in the experiment, so the experimental average, maximum and minimum values are also plotted (blue symbols). Again, these comparisons show that the predicted values are within or close to within the experimental range of values, and that the shape of the peak is similar between experimental and numerical. The numerical predictions capture the nature of the initial short duration peak, which is present in the no and half submergence cases.

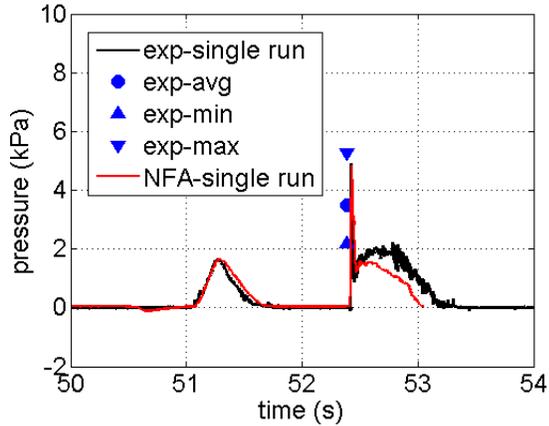

**Figure 20:** Comparison of experimental results for breaking wave from G2 and computational results at 18.10 cm (7.13 in) above the bottom of the cube with 0° cube angle and half submergence.

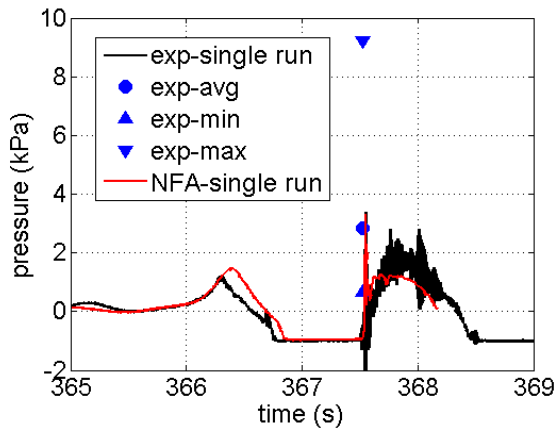

**Figure 21:** Comparison of experimental results for breaking wave from G11 and computational results at 4.13 cm (1.63 in) above the bottom of the cube with 0° cube angle and half submergence.

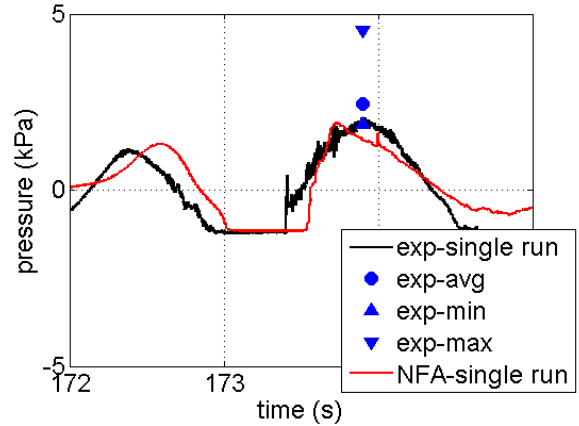

**Figure 22:** Comparison of experimental results for breaking wave from G2 and computational results at 18.10 cm (7.13 in) above the bottom of the cube with 0° cube angle and full submergence.

Figures 24-26 show a sequence of plots from the numerical simulations for the breaking wave impact on a cube with 0° angle and no submergence, half submergence, and full submergence respectively. The time of each snapshot is given in the lower left corner of each frame in seconds. The free surface is shown in blue and the pressure acting on the front and bottom of the cube are shown as color contours. The contour levels are provided at the top of each figure.

Figure 24 shows that the pressures on the bottom of the cube can be as great as or greater than the pressures acting on the front face. The greatest pressures occur at the stagnation points slightly behind from where jets of water form on the cube. The spray root on the bottom face is visible from t=2.6 sec to t=2.7 sec.

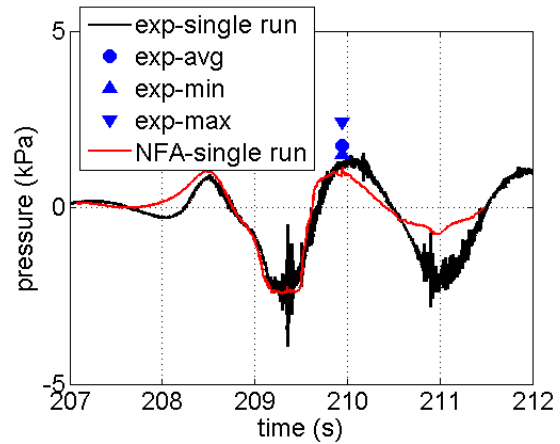

**Figure 23:** Comparison of experimental results for breaking wave from G11 and computational results at 4.13 cm (1.63 in) above the bottom of the cube with 0˚ cube angle and full submergence.

Figures 25 and 26 show that the time in which the cube is subjected to impact loading is longer for the full and half submergence cases compared to the zero submergence case.

**CONCLUSIONS**

The computational fluid dynamics code, Numerical Flow Analysis (NFA) has been used to predict wave impact pressures for incident non-breaking and breaking wave cases. Comparisons of the wave time series in front of the cube show that NFA generated a similarly shaped wave to that generated during the experiment. The predicted pressures are within or close to within the range of experimental values measured for both breaking and non-breaking waves and the shape of the peaks compare well. The ability of NFA to predict the breaking wave impact pressures is especially promising, since these tend to cause larger loads than the more predictable non-breaking wave loads. In terms of future research, we will use wavemaker data as input to ensure that the wave conditions for the experiments and numerics match exactly.

**ACKNOWLEDGEMENTS**

The Office of Naval Research supports this research. Dr. Ronald Joslin is the program manager. This research is also supported by a SAIC IR&D effort. Animations of NFA simulations are available at http://www.youtube.com/waveanimations. NFA predictions are supported in part by a grant of computer time from the DOD High Performance Computing Modernization Program. NFA simulations have been performed on the SGI Altix Ice at the U.S. Army Engineering Research and Development Center (ERDC).

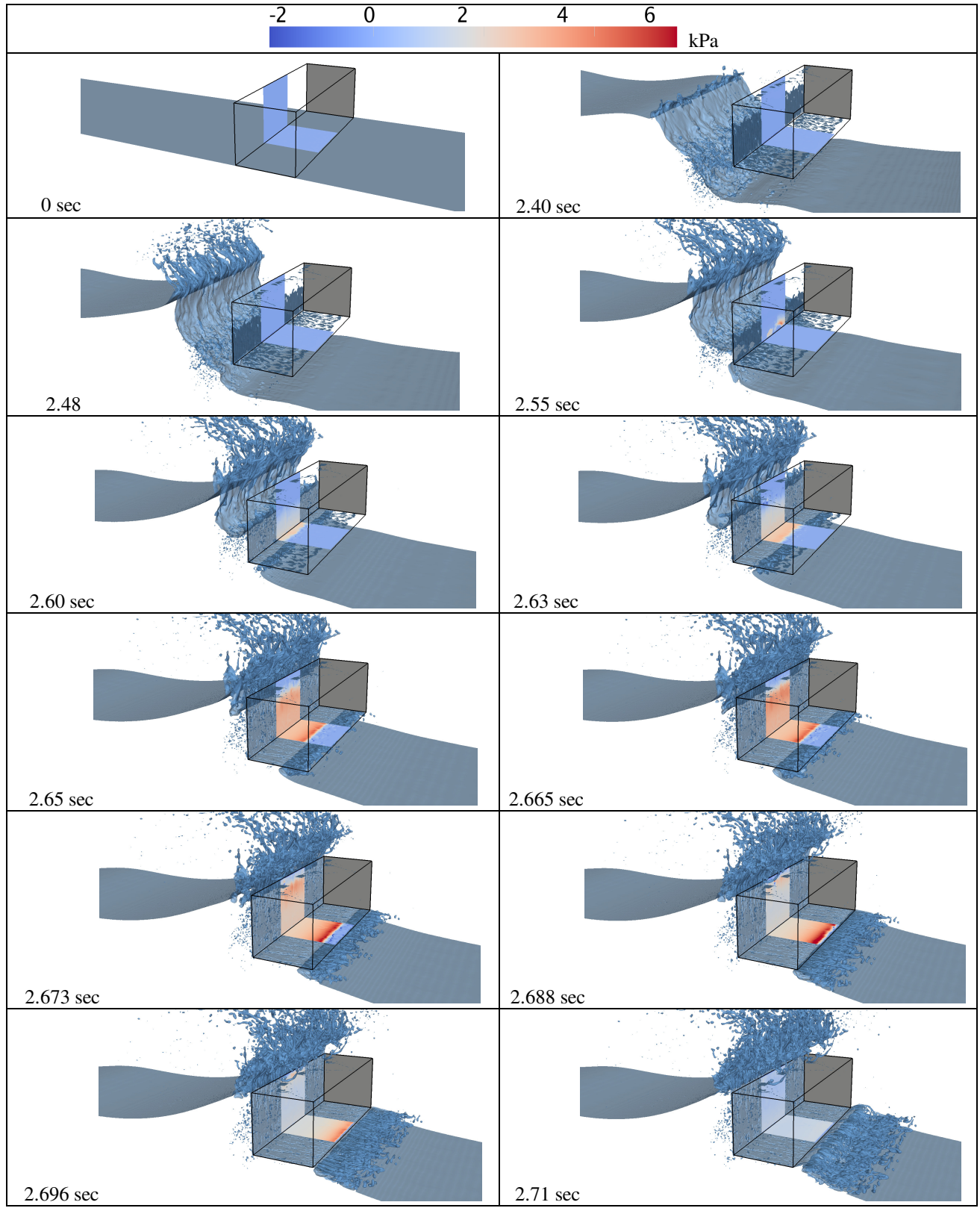

**Figure 24:** Time series of pressures for breaking wave, with cube above MWL.

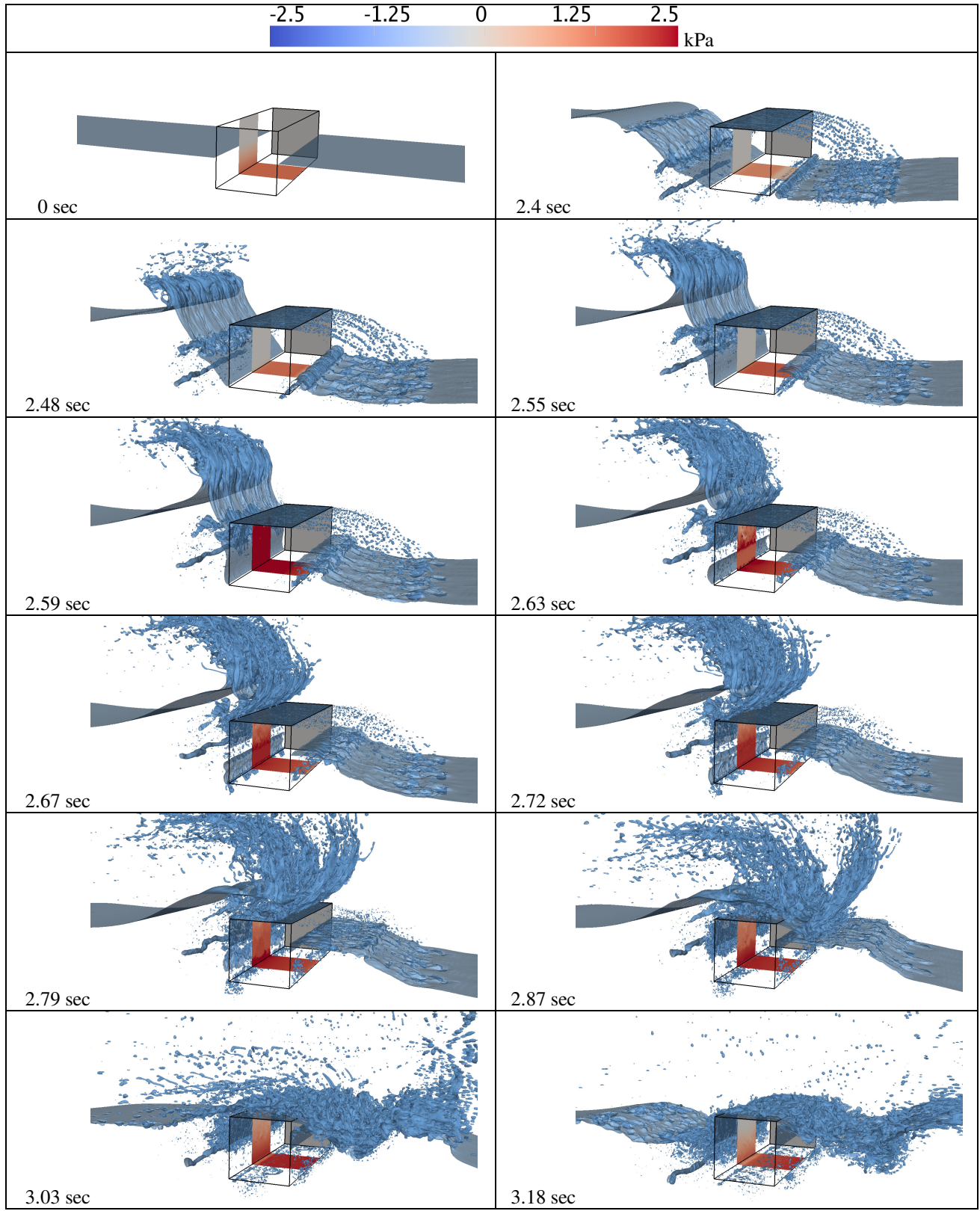

**Figure 25:** Time series of pressures for breaking wave, with cube half below MWL.

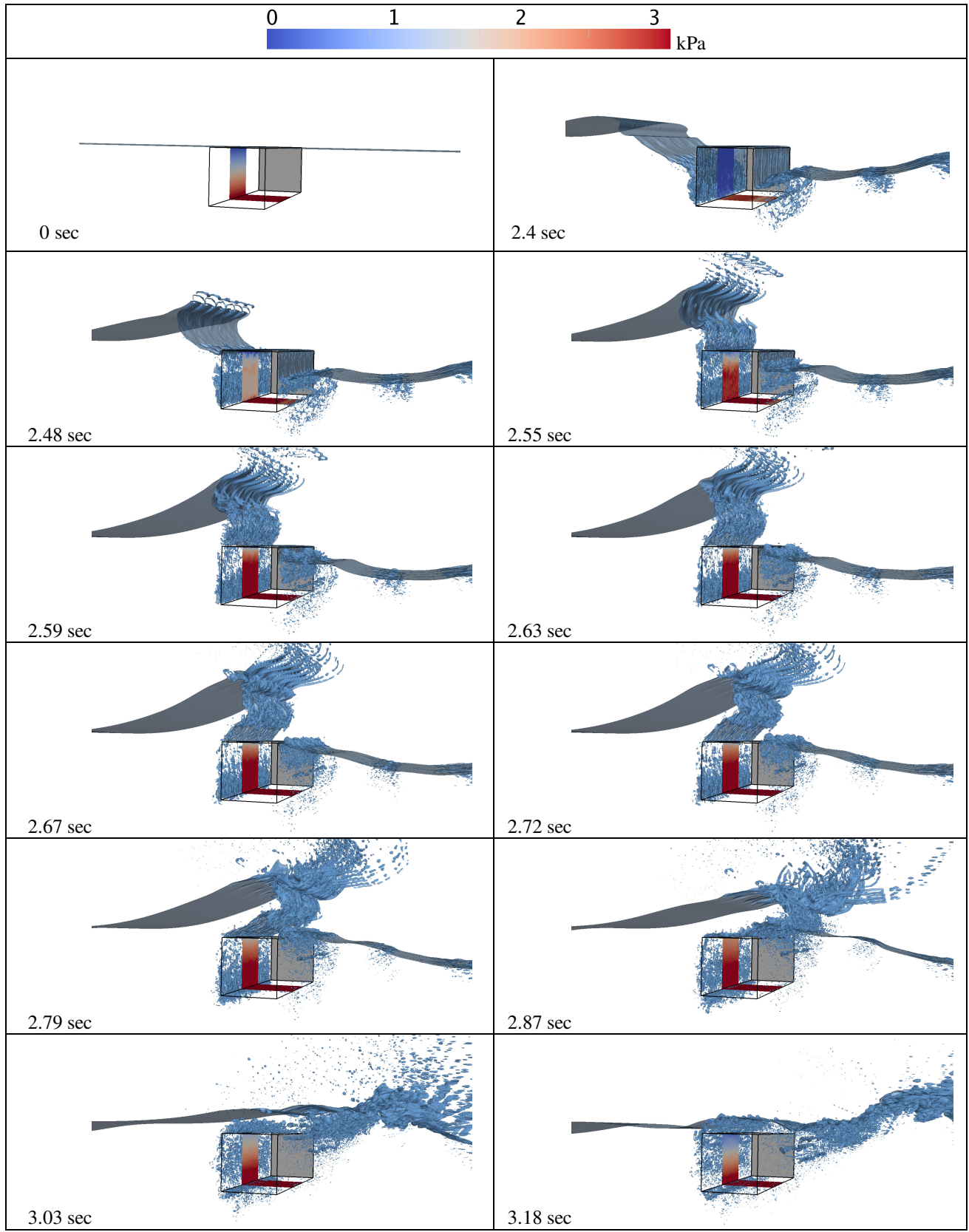

**Figure 26:** Time series of pressures for breaking wave, with cube below MWL.